\documentclass[fleqn,usenatbib]{mnras}

\usepackage{newtxtext,newtxmath}

\usepackage[T1]{fontenc}

\DeclareRobustCommand{\VAN}[3]{#2}
\let\VANthebibliography\thebibliography
\def\thebibliography{\DeclareRobustCommand{\VAN}[3]{##3}\VANthebibliography}

\usepackage{graphicx}	
\usepackage{amsmath}	
\usepackage[normalem]{ulem}
\usepackage{subcaption}
\author[M. J. Smith et al.]{
    Michael J. Smith,$^{1, 2, 3}$\thanks{E-mail: m.smith28@herts.ac.uk}
    Nikhil Arora,$^{3}$
    Connor Stone,$^{3}$
    St\'ephane Courteau,$^{3}$
    and James E.~Geach$^{1, 2}$
\\
    $^{1}$Centre for Astrophysics Research, Department of Physics, Astronomy \& Mathematics, University of Hertfordshire, Hatfield, AL10 9AB, UK\\
    $^{2}$Centre of Data Innovation Research,  Department of Physics, Astronomy \& Mathematics, University of Hertfordshire, Hatfield, AL10 9AB, UK\\
    $^{3}$Department of Physics, Engineering Physics \& Astronomy, Queen's University, Kingston, ON, K7L 3N6, Canada
}

\pubyear{2021}

\title[Pix2Prof]{Pix2Prof: fast extraction of sequential information from galaxy imagery via a deep natural language `captioning' model}

\begin{document}
\label{firstpage}
\pagerange{\pageref{firstpage}--\pageref{lastpage}}
\maketitle

\begin{abstract}
    We present `Pix2Prof', a deep learning model that can eliminate any manual
    steps taken when extracting galaxy profiles.  We argue that a galaxy profile
    of any sort is conceptually similar to a natural language image caption.
    This idea allows us to leverage image captioning methods from the field of
    natural language processing, and so we design Pix2Prof as a float sequence
    `captioning' model suitable for galaxy profile inference. We demonstrate
    the technique by approximating a galaxy surface brightness (SB) profile
    fitting method that contains several manual steps.  Pix2Prof processes
    $\sim$1 image per second on an Intel~Xeon~E5-2650~v3~CPU, improving on the
    speed of the manual interactive method by more than two orders of
    magnitude.  Crucially, Pix2Prof requires no manual interaction, and since
    galaxy profile estimation is an embarrassingly parallel problem, we can
    further increase the throughput by running many Pix2Prof instances
    simultaneously.  In perspective, Pix2Prof would take under an hour to infer
    profiles for $10^5$ galaxies on a single NVIDIA DGX-2 system. A single
    human expert would take approximately two years 
    to complete the same task.
    Automated methodology such as this will accelerate the analysis of the next
    generation of large area sky surveys expected to yield hundreds of millions
    of targets. In such instances, all manual approaches -- even those involving a
    large number of experts -- will be impractical.  
\end{abstract}



\section{Introduction} \label{sec.intro}

Large astrophysical surveys such as the Sloan Digital Sky Survey
\citep[SDSS;][]{sdss}, the Panoramic Survey Telescope and Rapid Response
System \citep[Pan-STARRS;][]{pan-starrs}, the Hyper Suprime Cam
\citep[HSC;][]{hsc} Subaru Strategic Program Survey, or the upcoming
Vera C.  Rubin Observatory Legacy Survey of Space and Time \citep[LSST;][]{lsst},
carry an inherent scaling problem.  The SDSS has observed over
35 per cent of the sky, cataloguing over 1 billion astronomical objects
\citep{sdss_dr16} with a data rate of raw multiband imagery approaching
200\,GB per night.  In January 2019, Pan-STARRS second Data Release totalled
1.6~PB of imaging data. A precursor to LSST, HSC's 990~megapixel camera has
already produced over 1~PB of imaging data \citep{hsc_dr2}. These surveys will
be dwarfed by the upcoming LSST project. LSST's 3.2~gigapixel camera will be
the largest ever made, and will survey the entire visible sky twice per week,
generating $\sim$500~PB of imaging data over its decade-long mission. The
`firehose' of data from surveys such as LSST will require correspondingly
efficient and fully automated procedures to curate and analyse the data,
enabling new astrophysical findings and making unanticipated discoveries. 

In this study, we are concerned with the automated direct analysis of galaxy
imagery towards estimating galaxy properties such as size, luminosity, colour,
and stellar mass. To calculate these properties, one typically applies a
photometric analysis that involves extracting and characterising the spatial
distribution of a galaxy's light, described by a surface brightness (SB)
profile. The galaxy structural parameters as reflected by the SB profile can be
used to infer a suite of other important characteristics such as light
concentration, age, star formation rate, and assembly history
\citep[e.g.][]{strom1976, bell2003, shen2003, bernardi2005, fernandez2013,
trujillo2020}.

Numerous catalogues of galaxy structural properties already exist
\citep{jedrzejewski1987, courteau1996, Brinchmann2004, Blanton2005, Hall2012,
Gilhuly2018}.  Unfortunately, the methods used in these compilations are either
time consuming, requiring human supervision, or fast but unreliable since they
require \textit{a priori} assumptions about the shapes of galaxy components and
other features. Similarly to the study detailed in this paper,
\citet{tuccillo2018} describe a fully automated neural network based technique
(named `DeepLeGATo'; Deep Learning Galaxy Analysis Tool) designed to
replicate the GALFIT model-dependent algorithm. DeepLeGATo is a fine example of
an effective application of deep learning, providing faster and possibly more
accurate analysis than its parent method, GALFIT \citep{GALFIT}.  However,
DeepLeGATo inherits from its similarity to GALFIT the same issues stated
previously; namely, the hard-coded assumption of galaxy profile shapes, and
other features.  Furthermore, DeepLeGATo can only produce single float outputs,
and so cannot infer an SB profile directly. This means that DeepLeGATo must rely
on an intermediary model to generate a complete SB profile.  Therefore, even
with semi-automated methods, the accurate extraction of {\it all} the useful
information from existing surveys would take years.  With the data volume
expected to grow significantly in the coming years, this becomes an intractable
problem.  Of great concern is the possibility that important discoveries and
insight could be missed or delayed significantly due to the technical
challenges imposed by the unprecedented data volume.  Clearly, there is a
pressing need for entirely new and efficient automated methods that
significantly reduce, and ideally circumvent, human interactions.  Machine
learning is ideally suited for this task, and we apply it in this paper towards
the specific problem of extracting SB profiles from multi-band imaging data.
Our approach takes advantage of a set of SB profiles already determined via
classical, interactive methods \citep{courteau1996,Gilhuly2018}.  We describe
the classical method used to produce the training data set in the next section.
The remainder of the paper is organised as follows: Section \ref{sec.method}
introduces our approach; our results and validation are presented in Section
\ref{sec.results}; Section \ref{sec.dandc} addresses our global findings, and
concludes with suggestions for broader application of the algorithm.

\section{Method} \label{sec.method}

\subsection{The classical surface brightness profile extraction algorithm} \label{sec.courteau}

In the surface photometry of galaxies \citep[e.g.][and references
therein]{courteau1996}, the spatially resolved light profile of a galaxy is
extracted by fitting progressively larger isophotes about a common centre.  The
fitting technique assumes that projected isophotes are well represented by
ellipses. A galaxy's centre is found by identifying the brightest pixel in a
manually selected region. Given a manually defined galaxy centre, the classical
algorithm determines the parameters needed to define each ellipse via a least
squares optimisation. The algorithm then generates isophotal solutions at each
radius well into the faint outskirts of the galaxy.  In these regions of lower
signal-to-noise, where fitting algorithms are challenged, the algorithm
radially extends the last fitted isophote in the previous operation with a set
of concentric isophotes out to an arbitrarily large radius, usually taken to be
the edge of the image.

The isophotes may vary in ellipticity and position angle as a function of
galactocentric radius.  This can become problematic when fitting to
non-axisymmetric structures in galaxies, such as bars and spiral arms that can
cause large twists in the fitted isophotal map.  This issue can be corrected by
manually applying a smoothing function to some portions of the image. The
latter consists of manually smoothing the contour fits (i.e.\ uncrossing
twisted isophotes), and replacing poorly fitted data with a polynomial
smoothing function.  Note that, prior to applying isophotal fitting to galaxy
images, some pre-processing is also required: the galaxy centre must be
identified as described above; the `sky' background must be estimated and
removed from the image; nuisance foreground objects (such as unassociated
galaxies or foreground stars) must be identified and masked.  These steps add
to the manual supervision of the task.

Besides the assumption that galaxies are circular when viewed face-on, and thus
generally of elliptical appearance when projected on to the plane of the sky,
the algorithm purposefully avoids using \textit{a priori} knowledge of galactic
disc profile shapes and other features such as bars, rings, and spiral arms.
This avoids biasing the isophotal solution to any pre-determined, and possibly
incorrect, shape -- a problem especially acute in the faint outer edges of a
galaxy.

\begin{table}
 \caption{A summary of the \citet{courteau1996} surface brightness profile
    fitting algorithm's processes. An approximate wall time per galaxy is given
    for the manual sections.  The automated sections' time contributions are
    negligible.}
  \centering
  \begin{tabular}{l c c}
  \hline 
  Process & Automated? & Wall time (s/gal) \\
  \hline
  Identify galaxy centre &              No  & 5 \\
  Estimate \& remove sky background &   Yes & -- \\
  Remove foreground objects &           No  & 120 \\
  Fit contours &                        Yes & -- \\
  Extend contours to galaxy extent &    No  & 30 \\
  Smooth isophotes &                    Yes & -- \\
  Interpolate poorly fitted data &      No  & 120  \\
  \hline
  \end{tabular}
  \label{tab.block}
\end{table}

While the semi-automated steps outlined above yield high quality SB profiles,
the process of obtaining a single profile is slow and systematic
variations may exist between different profile extraction methods. The
interactive nature of certain steps may indeed give rise to marked profile
differences, especially in low SB regimes where the isophotal solutions are
less robust. The low SB regimes will always remain the bane of galaxy image
analysis, whether automated or interactive, but the elimination of subjective
steps goes a long way towards reducing systematic differences between profiles.
Therefore, it becomes desirable to eliminate all interactive steps while
retaining all the benefits of classic algorithms such as \citet{courteau1996}
described above. In this work, we present a fully automated solution that
incorporates the extant knowledge base of SB profile fitting methodology, 
but avoids human interaction. 

\subsection{Borrowing from automated image captioning}

In recent years, the field of automated image captioning has benefited greatly
from advances in deep learning.  We were strongly influenced by these
developments when designing the architecture of our `Pix2Prof' profile
estimator.  In this section, we briefly review pure recurrent neural network
(RNN) based encoder--decoder architectures, or models that only use a
single encoder and decoder to generate captions.  A comprehensive review on
deep learning methods applied to image captioning can be found in
\citet{captioning_review}.

We primarily draw inspiration from gated RNN based encoder--decoder
architectures, as seen in \citet{seq2seq} for sequence-to-sequence translation,
and in \citet{showandtell}, \citet{gLSTM}, and \citet{bidirectional_captioning}
for image-to-sequence translation. \citet{seq2seq} uses a Long Short Term
Memory \citep[LSTM;][]{LSTM} network to encode a given sentence to a
latent descriptive vector, and a second LSTM network to decode the descriptive
vector into a different feature space. One can use this technique to translate
text between two different languages, for example.

\begin{figure*}
    \centering
    \includegraphics[width=\textwidth]{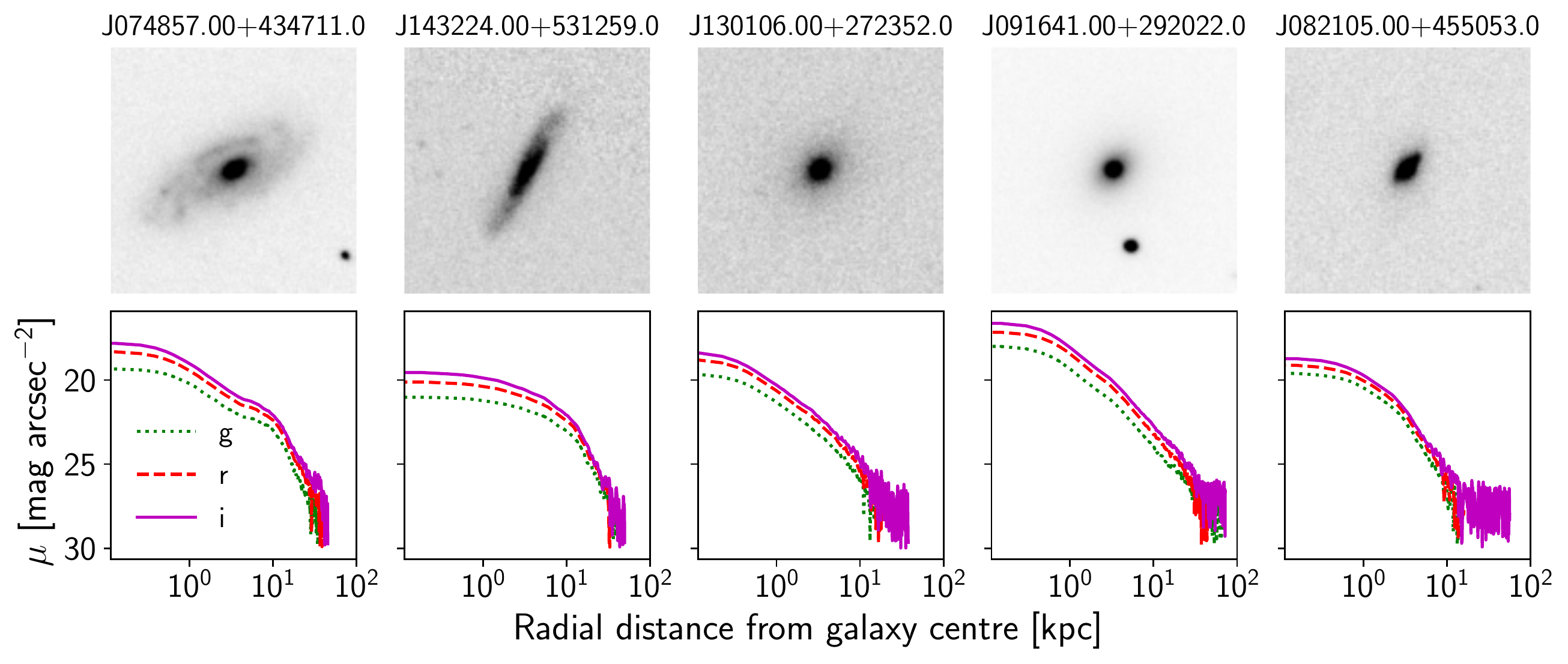}
    \caption{SDSS images of sample galaxies in the $g$ band (top row),
        and corresponding surface brightness `ground truth' profiles (bottom
        row). $\mu$ is the surface brightness.  The galaxy names above each
        image refer to their J2000 celestial coordinate.}
    \label{fig.training_set} 
\end{figure*}

\citet{showandtell}, \citet{gLSTM}, and \citet{bidirectional_captioning} all
use a convolutional neural network \citep[CNN;][]{neocognitron,
lecun1989} to first encode an image to a latent descriptive vector, and then
use an LSTM network to decode this vector into a text description (caption) of
a given image. \citet{showattendandtell} use a CNN encoder, and a LSTM that
attends over the CNN output. Attention allows this approach to link each word
in the caption with an associated part of the image. This approach works well
for images that are crowded with multiple objects, but a simpler approach is
preferred for our case where each image is dominated by a single, central
galaxy.

A galaxy profile can be thought of as analogous to a text caption describing
that galaxy. Both a text caption and galaxy profile can be encoded as a list of
floats or integers, and both have a length and content dependent on the context
of the conditioning image. Both also need to terminate once a complete sentence
or profile is generated, a subjective task well suited to a machine learning
solution.  RNNs learn where to terminate a given caption or profile empirically
from the training set.  Additionally, galaxy profiles and text captions can
both be approximated with an appropriate RNN.  RNNs also produce spatially
consistent captions as a consequence of the architecture, a property not
guaranteed in a straight one-shot CNN.  With these benefits in mind, it is
natural to consider an encoder--decoder network for the specific task of
estimating challenging galaxy profiles. Importantly, since the proposed model
directly learns the transformation between an unprocessed galaxy image, and the
galaxy's corresponding SB profile, it eliminates all of the manual steps
described in Section~\ref{sec.courteau} and Table~\ref{tab.block}.

While we develop Pix2Prof within the context of galaxy SB profile extraction,
the model is equally applicable to any array $\rightarrow$ float sequence
translation task.  

\subsection{Training set} \label{sec.training set}

We initially populate our data set with 10 arcmin $\times$ 10 arcmin $g$, $r$,
and $i$ band image crops extracted via the SDSS DR10
\citep{sdss,sdss_iii,sdss_dr10} online mosaic interface. Each image is centred
on a galaxy. From these images we extract an SB profile via the method described
in Section~\ref{sec.courteau}.  Fig.~\ref{fig.training_set} presents a random
sample of training set galaxies, and their corresponding, manually extracted SB
profiles.  The 1953 galaxy image--SB profile pairs in each of the $g$, $r$, and
$i$ bands yield a total of 5859 pairs.  This full data set is then divided into
training, validation, and testing sets.  There are 5367 galaxy image--profile
pairs in the training set, 192 galaxy image--profile pairs in the validation
set, and 300 galaxy image--profile pairs in the test set.  The sets are
randomly assigned, with the condition that a given galaxy's three photometric
bands are kept within the same set.  The subset sizes are chosen to maximise
the training set efficacy while retaining most of the training set variance in
the test set.

The only destructive pre-processing performed on the galaxy imagery is a 99.9th
percentile clipping.  This clipping mitigates the issue of single bright (i.e.\
`hot') pixels reducing image contrast when the galaxy images are normalised, which
would reduce training efficacy.  To this end, we apply a fixed min-max
normalisation, defined as
\begin{equation}
    \bar x = \frac{x - A}{B - A},
\end{equation}
where $A = 2.0~\text{nanomaggies}$ is the floor of the minimum value in the
training set, and $B = 30.0~\text{nanomaggies}$ is the ceiling of the 99.9th
percentile value in the training set.

To reduce information sparsity in the training set images, we further
crop each galaxy image to a shape of \texttt{[256, 256]} pixels before
passing the image to Pix2Prof. We train using the full 32-bit depth of the
original data as measured.  Good quality data are paramount when training a
neural network, and we therefore cut the profiles when the signal-to-noise
ratio reaches a quality threshold.  We terminate the profile when the
signal-to-noise ratio of a 1D convolution with a length of 40 pixels
($=8 ~\text{arcsec}$) reaches a threshold of 4.  We define signal-to-noise so
that it is equivalent to the ratio of the power of a signal to the power of
background noise: $(\mu/\sigma)^2$, where $\mu$ is the mean and $\sigma$ is the
standard deviation of the convolutional window. 

\subsection{Network architecture} \label{sec.arch}

\begin{figure*}
    \centering
    \includegraphics{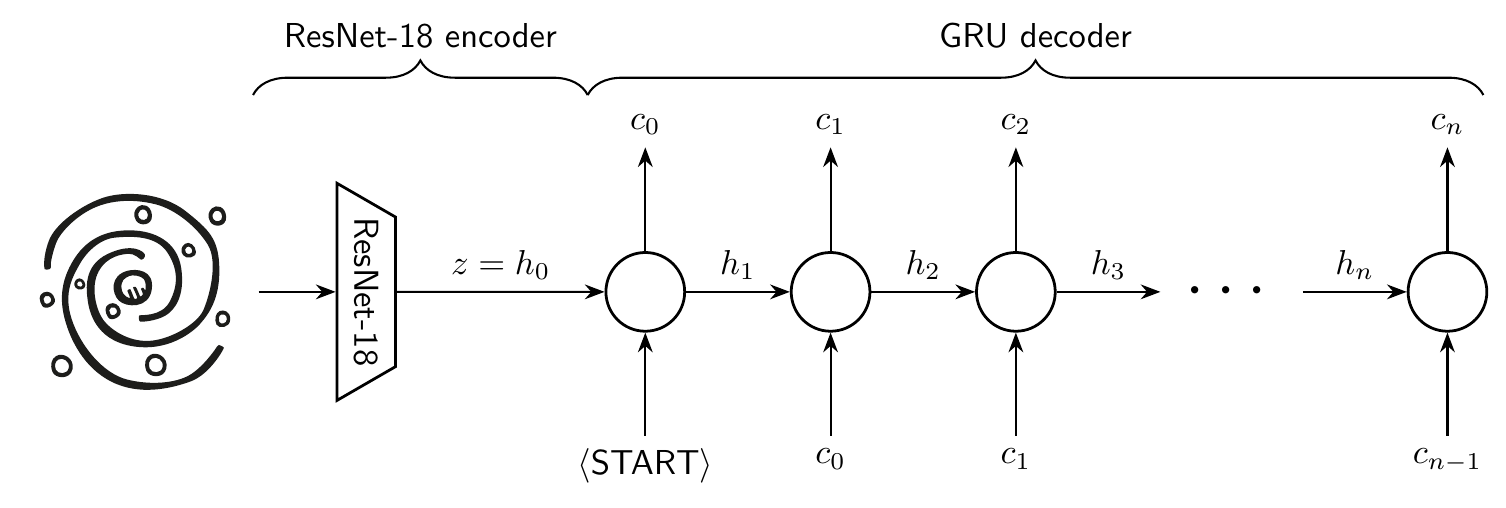}
    \caption{The ResNet $\rightarrow$ GRU encoder--decoder architecture used in
        this work. The hidden state $h_i$ is the internal state of the GRU, and is
        dependent on both the galaxy latent encoding $z$, and the previous profile
        predictions $c_i$.}
    \label{fig.arch}
\end{figure*}

We write our model in PyTorch \citep{pytorch}, using a ResNet-18 \citep{highway,resnet} 
encoder, and a Gated Recurrent Unit \citep[GRU;][]{GRU} decoder
architecture.  This architecture takes an arbitrarily sized single channel
image input, and outputs a sequence of floats of arbitrary length.
These floats are spaced along a galaxy's semimajor axis at a spacing of 0.2
arcsec per step (the same as the \citet{courteau1996} technique).
The same network is trained on images in the $g$, $r$, and $i$ bands, and
therefore can produce a SB profile in any one of these bands.
Fig.~\ref{fig.arch} shows a representation of the architecture used. 

We use the standard ResNet-18 architecture as described in \citet{resnet}.  The
GRU is stacked to three layers.  We apply a rectified linear unit
\citep[ReLU]{relu} activation and a dense neural layer after the
three layer stack to reduce the number of output values to one.  As a
regularising measure, we apply dropout at a 20 per cent rate \citep{dropout}.  The
ResNet first encodes the incoming galaxy image to a latent space vector $z$ of
length $512$.  This vector is then used as the initial hidden state $h_0$ of a
GRU. In this way, Pix2Prof encodes and passes relevant information from the
image to the GRU. The GRU then unrolls to estimate properties of the galaxy
from $z$.  In this paper's case, we demonstrate this process by using $z$ to
estimate a galaxy's SB profile.

To start estimation, the GRU is fed a start of sequence token. This token is
set as an array of zeros. In place of an end of sequence token, the GRU is
programmed to halt after 100 predictions are output that have a standard
deviation of 0.01 or less.  This ensures that the GRU halts estimation once it
encounters the background sky. 

We use the Adam optimiser \citep{adam} to train Pix2Prof via gradient descent
\citep{grad_descent}. Using manual search, we set the learning rate as $2
\times 10^{-4}$.  Due to the logarithmic nature of magnitude, we want to
penalise large deviations from our ground truth SB profiles at a higher rate
compared to small deviations, and so we use the mean squared error loss: 
\begin{equation} 
    \text{MSE} = \frac{1}{b} \sum^b_{i = 1} (y_i - p_i)^2,
\end{equation}
where $b$ is the batch size, $y$ is the ground truth, and $p$ is a prediction.

\subsection{Training the model} \label{sec.training}

We augment the galaxy images by applying a `wobble'.  This wobble is a random
small shift in the centre of the image. Each band is treated independently.  We
do this to encourage the network to work with the slightly off-centre galaxies
that will be encountered in real data.  This is required to make Pix2Prof
robust to poorly centred galaxy images. We also exploit the rotational
axisymmetry of galaxies and further augment the data by randomly rotating an
input image through $90, 180, \text{and}~270$ degrees.

We train the model for $100\,000$ global steps on a single NVIDIA TESLA V100 GPU.
Training takes approximately 20~minutes per epoch of 500 galaxy images, a rate
of 0.4 galaxies per second.

\section{Results \& Validation} \label{sec.results}

We validate the model during training once per epoch using the validation set.
We test the trained model on 100 randomly sampled observed galaxies in the $g$,
$r$, and $i$ bands (for 300 total image--profile pairs) drawn from the data set
and which are set aside entirely during training.  We use the model with the lowest
validation loss; epoch 160.  We run an entirely automated inference on an
Intel~Xeon~CPU~E5-2650~v3~CPU at a rate of $0.9$ galaxies per second.

\begin{figure*}
    \centering
    \includegraphics[width=\textwidth]{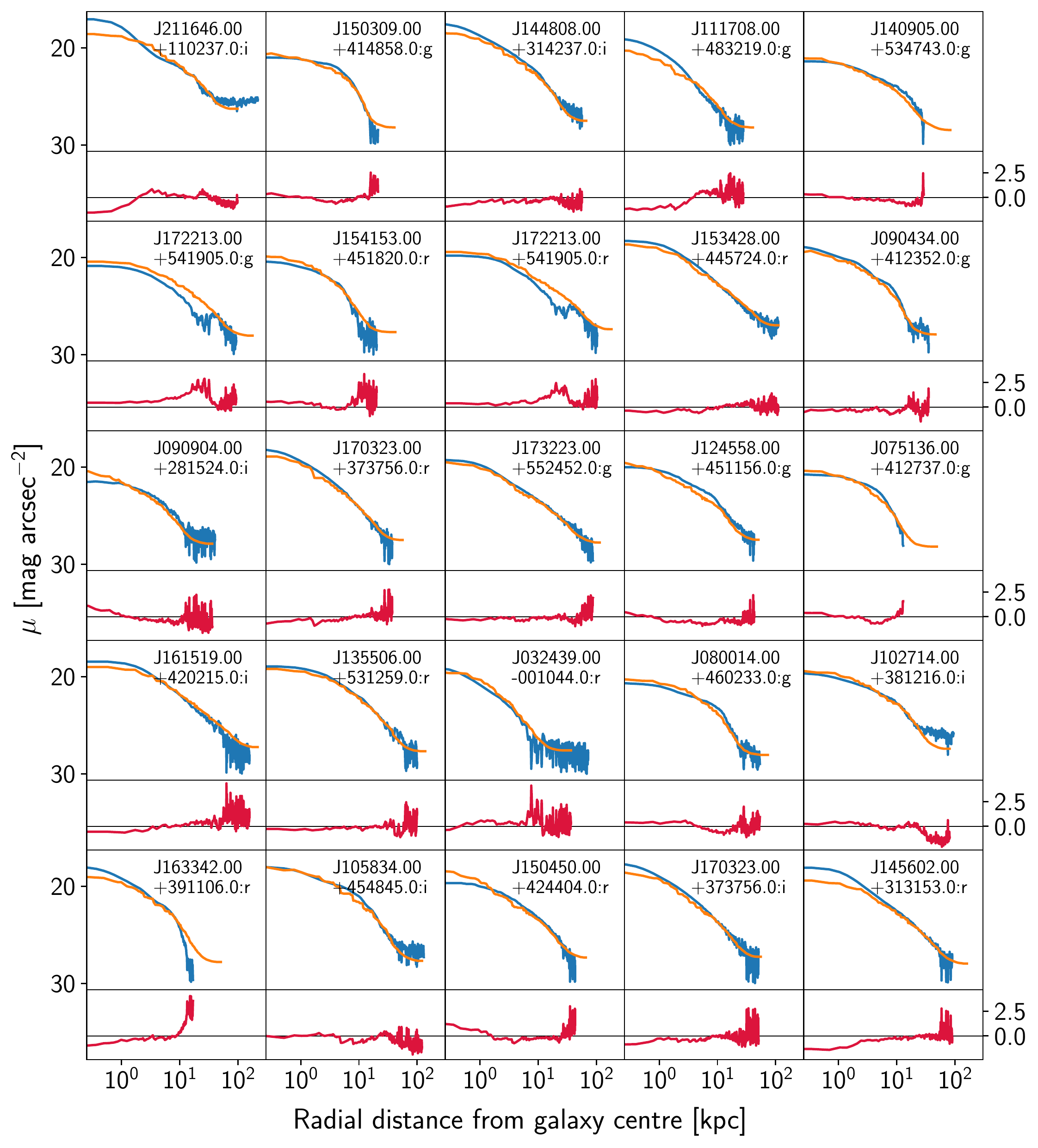}
    \caption{Randomly sampled test set predicted SB galaxy profiles (orange)
        superimposed onto SB profiles measured via the \citet{courteau1996} method
        (blue). $\mu$ is the surface brightness. Distances from centre are in log scale
        to emphasise divergences in the high signal-to-noise ratio region closer to
        the galaxies' centres.  Below each SB profile plot is the residual defined
        as $\text{res} = y - p$, where $y$ is the profile as measured according to
        Section~\ref{sec.courteau}, and $p$ is the prediction. The galaxies' J2000
        celestial coordinates and spectral bands are indicated at the top right
        of each graph.}
    \label{fig.results}
\end{figure*}

\begin{figure*}
    \centering
    \begin{subfigure}[t]{\textwidth}
        \centering
        \includegraphics[width=\textwidth]{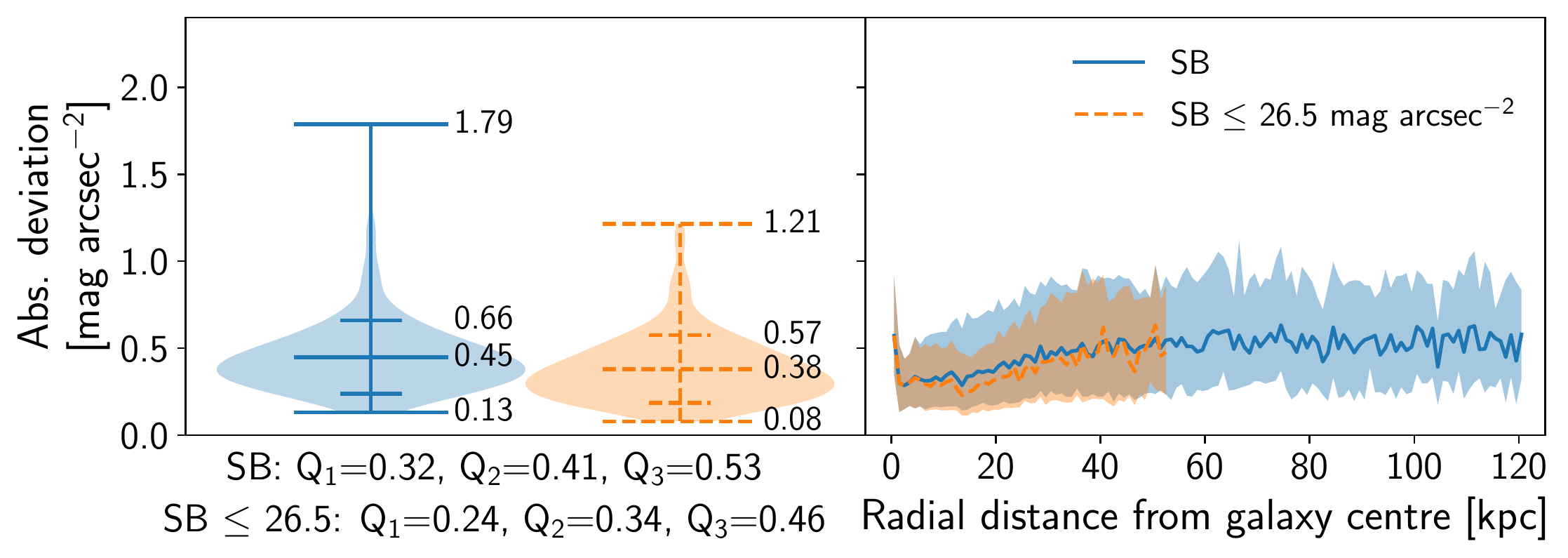}
        \caption{Summary statistics for Pix2Prof.} 
        \label{fig.pix2prof_stats}
    \end{subfigure}
    
    \begin{subfigure}[t]{\textwidth}
        \centering
        \includegraphics[width=\textwidth]{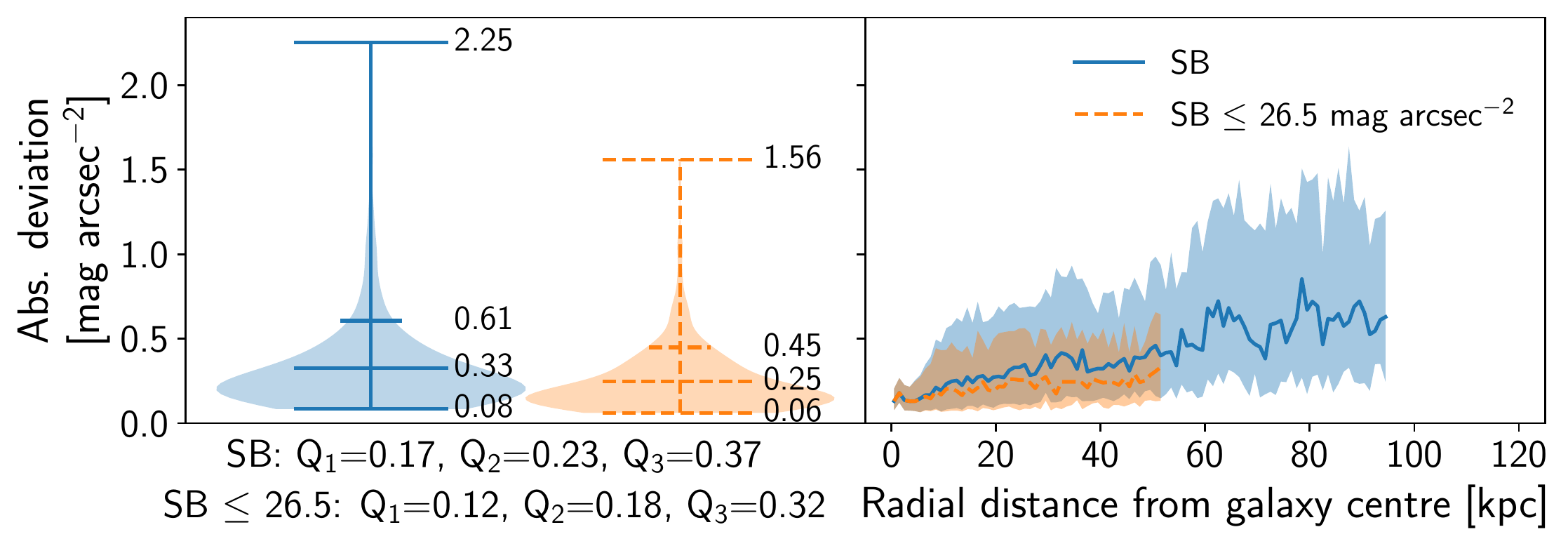}
        \caption{Summary statistics for AutoProf (see \S\ref{sec.autoprof}).}
        \label{fig.autoprof_stats}
    \end{subfigure}
    
    \begin{subfigure}[t]{\textwidth}
        \centering
        \includegraphics[width=\textwidth]{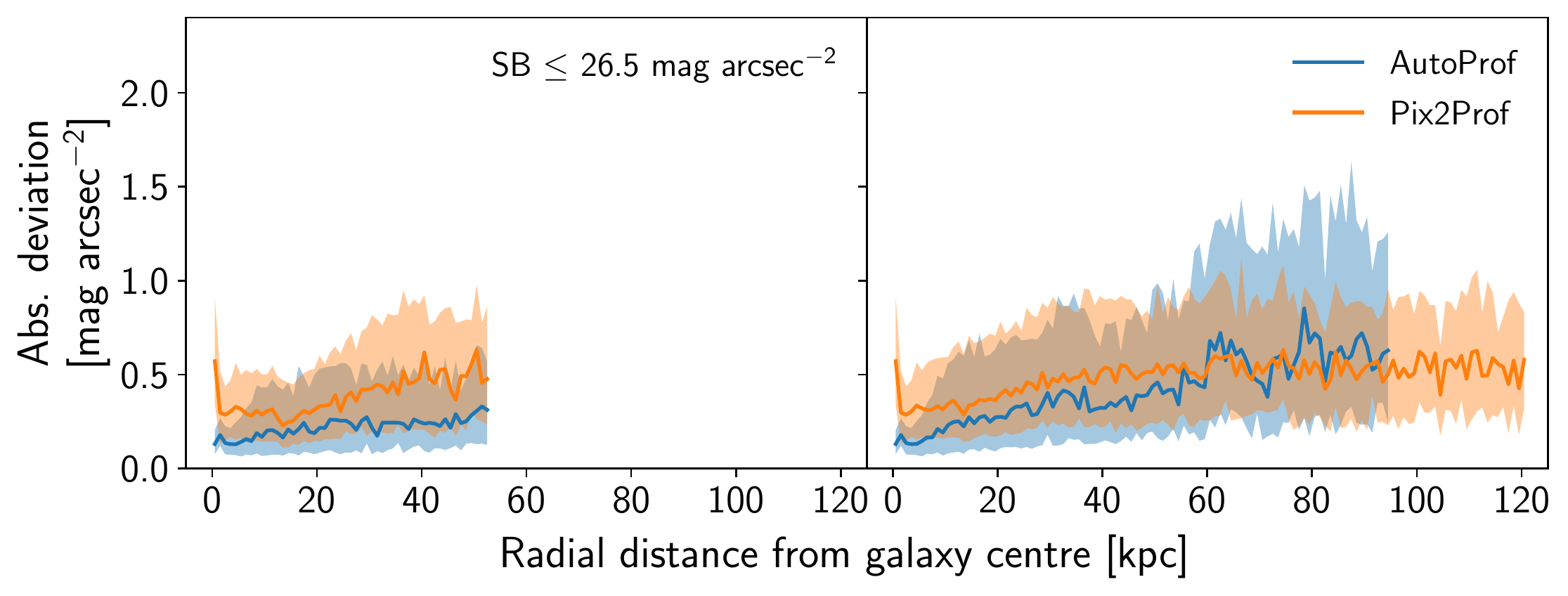}
        \caption{Median error per kpc from galaxy centre comparison between Pix2Prof and AutoProf.}
        \label{fig.autoprof_vs_pix2prof}
    \end{subfigure}
    \caption{Approximation errors as defined in equation~(\ref{eqn.err}).
        Fig.~\ref{fig.pix2prof_stats} shows summary statistics for Pix2Prof.
        For comparison, Fig.~\ref{fig.autoprof_stats} depicts the same
        statistics for AutoProf.  In all of the above figures we define
        absolute deviation as relative to the \citet{courteau1996} test set
        profiles.  The leftmost violin plot in Fig.~\ref{fig.pix2prof_stats}
        and \ref{fig.autoprof_stats} shows the distribution of median test set
        errors.  The rightmost violin plot in Fig.~\ref{fig.pix2prof_stats}
        and \ref{fig.autoprof_stats} shows the same distribution for only SB
        values below the SDSS limiting SB of $26.5$~mag~arcsec$^{-2}$. The
        maximum, minimum, mean, and (mean + standard deviation) are labelled.
        Below the violin plots are their distribution quartiles.  The line
        plots show the median error per kpc from the galaxy centre, with the
        interquartile range shaded. To reduce the effect of small sample size
        variability, the line plots are terminated once 90 per cent of the SB profiles
        reach their galaxy's extent.  Fig.~\ref{fig.autoprof_vs_pix2prof}
        compares on the same axes the median errors per kpc from the galaxy
        centre for Pix2Prof and AutoProf.}
    \label{fig.errors}
\end{figure*}

Fig.~\ref{fig.results} shows a random selection of 25 Pix2Prof inferred test
set SB profiles superimposed on to the \citet{courteau1996} SB profiles.  Since
we have trained Pix2Prof to directly infer a SB profile from a galaxy image we
do not produce intermediate steps (such as the galaxy centre, ellipticity
profile, or position angle profile).  However, one could estimate these values
if Pix2Prof is explicitly trained to reproduce them. Fig.~\ref{fig.errors}
shows the error distribution of the test set as well as the test set error per
distance in physical units from the galaxy centre. We define error as the
absolute of Fig.~\ref{fig.results}'s residual, the absolute deviation: 
\begin{equation} \eta = \left|y - p\right|, \label{eqn.err} \end{equation}
where $p$ is a prediction, and $y$ is measured via \citet{courteau1996}.  The
units of SB call for additional care in defining our errors. Since SB values
are defined on a logarithmic scale, equation~(\ref{eqn.err}) is really a form of
fractional error:
\begin{equation} \eta = \left|2.5 \log_{10} \frac{I_p}{C} - 2.5 \log_{10}
\frac{I_y}{C}\right| = \left|-2.5 \log_{10} \frac{I_y}{I_p}\right|,
\label{eqn.frac_err} \end{equation}
where $C$ is a constant reference brightness. $\{I_p,I_y\}$ are brightnesses in
linear units.

We take the median of this error per galaxy profile to produce the violin plots
in Fig.~\ref{fig.pix2prof_stats}, and we take the median of this error across
profiles to produce the line plot in  Fig.~\ref{fig.pix2prof_stats}.
Fig.~\ref{fig.pix2prof_stats}'s line plot shows that the error increases with
radius away from the galaxy centre towards regions containing less signal, as
expected.  We find that the median test set absolute deviation is
0.41~mag~arcsec$^{-2}$ with an interquartile range of 0.21~mag~arcsec$^{-2}$.
We also find that the median test set absolute deviation for $y$ values
brighter than the SDSS limiting SB (26.5~mag~arcsec$^{-2}$) is
0.34~mag~arcsec$^{-2}$, with an interquartile range of 0.22~mag~arcsec$^{-2}$.
Errors of this scale mean that profiles generated via Pix2Prof will be
immediately useful for rough searches; it would be possible to categorise
galaxies roughly by brightness, isophotal radius, scale length, and other
structural parameters.  Further refinement of the model may reduce error,
enabling more sophisticated processing and analysis of generated SB profiles.
Possible refinements are described in Section~\ref{sec.dandc}.

\begin{figure*}
    \centering
    \includegraphics[width=\textwidth]{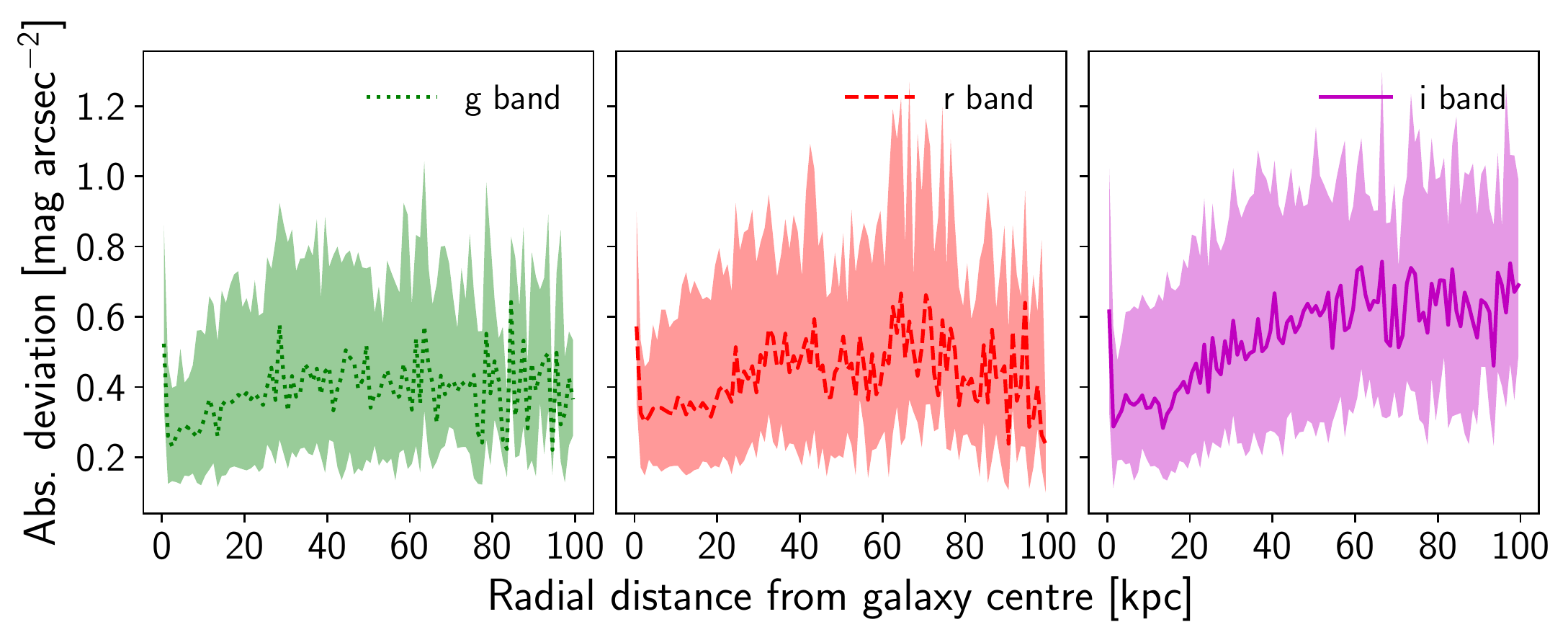}
    \caption{Median test set error per kpc from the galaxy centre, with the
        interquartile range shaded, split into the three bands present in the
        test set.}
    \label{fig.channelerrors}
\end{figure*}

In Fig.~\ref{fig.channelerrors}, the three bands' median errors are separated as
a function of galactocentric radius. Close to the galaxy, there is
little difference in the three bands' median predictions. However, as we proceed
outwards, the $r$-band's error is higher than the $g$-band's, and the
$i$-band's error is higher still. This is due to a difference in the
instrumental noise between the three bands, as evidenced in the difference in
the spectral bands' median galaxy image signal-to-noise ratios: $\text{SNR}_g =
41.6$; $\text{SNR}_r = 35.8$; $\text{SNR}_i = 28.4$.

\begin{figure*}
    \centering
    \includegraphics[width=\textwidth]{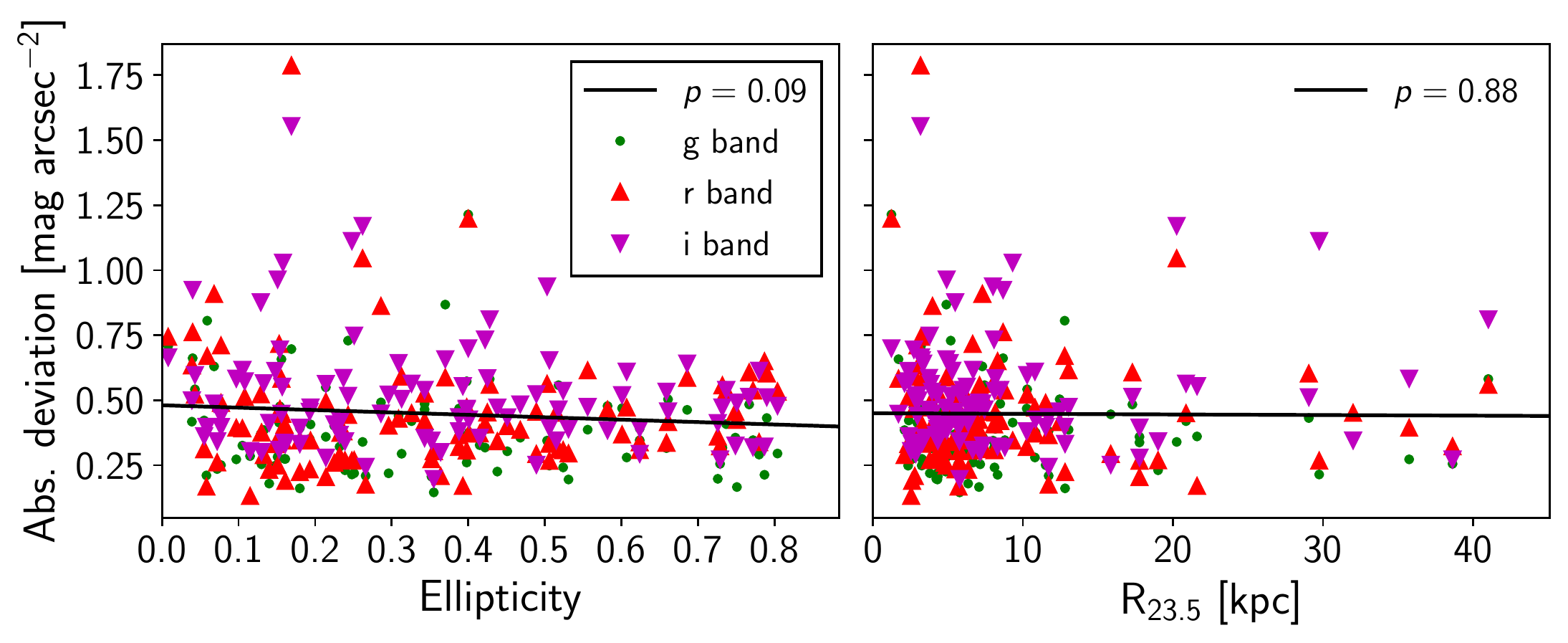}
    \caption{Median test set error over each galaxy predicted profile,
        plotted against each galaxy's ellipticity (left, \ref{fig.elliperr}a), and
        size at $R_{23.5}$ (right, \ref{fig.elliperr}b). $p$ values obtained via a
        linear regression are stated in the legends.}
    \label{fig.elliperr}
\end{figure*}

Fig.~\ref{fig.elliperr}a shows each test set galaxy's
ellipticity against the galaxy profile's median absolute deviation. The
ellipticity is defined as the final isophote ellipticity for a profile
calculated using the \citet{courteau1996} method.  Fig.~\ref{fig.elliperr}b
shows each test set galaxy's semimajor axis radius for the first isophote
whose value is greater than or equal to $23.5$~mag~arcsec$^{-2}$.  In both of
these cases, we run a linear regression and find no significant correlation,
suggesting that Pix2Prof's predictions are equally robust when inferring
across galaxies with a range of sizes and ellipticities.

Figs~\ref{fig.results} and \ref{fig.errors} show that Pix2Prof can
successfully approximate a complicated astrophysical image processing pipeline
with low deviation (0.34~mag~arcsec$^{-2}$ averaged over the test set).
Processing $0.9$ galaxies per second on an Intel~Xeon~E5-2650~v3~CPU, 
Pix2Prof improves on the speed of the classical image analysis method of
\citet{courteau1996} by more than two orders of magnitude. For comparison, 
an astronomer trained to use the \citet{courteau1996} method can typically 
process ${\sim}150$ galaxies in a full eight hour working day (or ${\sim}0.005$
galaxies per second). However, even astronomers must rest and so the true
working rate for a human would be ${\sim}150$ galaxies per 24 hours, 
or ${\sim}0.002$ galaxies per second.

As Table~\ref{tab.compar} shows, Pix2Prof eliminates any manual interaction
from SB profile inference, alleviating the issue of subjectivity in the
different methods developed for such tasks; Pix2Prof will infer the same
profile every time for a given galaxy image, whereas a human may not.  
The full automation of Pix2Prof enables a complete parallelisation, and thus 
significant gain in parallel throughput of galaxy profile estimation. 

\begin{table}
  \caption{Pix2Prof eliminates all interactive steps in the
      \citet{courteau1996} algorithm, alleviating subjectivity
      and accelerating inference significantly.}
  \centering
  \begin{tabular}{l c c}
  \hline 
          & \multicolumn{2}{c}{Automated in:} \\
  Process & \citet{courteau1996}? & \textcolor{red}{Pix2Prof}? \\
  \hline
  Identify galaxy centre &              No   & \textcolor{red}{Yes} \\
  Estimate \& remove sky background &   Yes  & \textcolor{red}{Yes} \\
  Remove foreground objects &           No   & \textcolor{red}{Yes} \\
  Fit contours &                        Yes  & \textcolor{red}{Yes} \\
  Extend contours to galaxy outskirts & No   & \textcolor{red}{Yes} \\
  Smooth isophotes &                    Yes  & \textcolor{red}{Yes} \\
  Interpolate poorly fitted data &      No   & \textcolor{red}{Yes} \\
  \hline
  \end{tabular}
  \label{tab.compar}
\end{table}

\subsection{Comparison with AutoProf} \label{sec.autoprof}

AutoProf \citep{stone2021} is a sibling method to Pix2Prof
that uses a more standard astronomical pipeline to tackle the problem of
non-parametric automated SB profile inference.  A combination of standard image
analysis packages from \textsc{photutils}~\citep{photutils} and novel techniques are
used to construct a robust isophotal pipeline.  Initial image analysis such as
determining the PSF, centre finding, star masking, and sky background
subtraction are performed using \textsc{photutils}.  Next, AutoProf simultaneously fits
an ellipticity and position angle profile by minimising low order FFT
coefficients along each isophote plus a regularisation term~\citep{shalev2014}.
The regularisation term penalises neighbouring ellipticity and position angle
values that deviate significantly, ultimately favouring smooth profiles.  Once
the profiles have converged, AutoProf extracts the median SB along each
isophote and determines the error with a 68.3 per cent quartile range.  A curve of
growth is determined by appropriately integrating the SB profile and
propagating errors.

Fig.~\ref{fig.errors} compares the performance of Pix2Prof and AutoProf.
AutoProf is found to produce SB profiles that are a slightly closer match to
those produced via \citet{courteau1996} than Pix2Prof. The difference in
absolute deviation between the two methods' profiles is typically around
0.1--0.2~mag~arcsec$^{-2}$.  However, AutoProf's accuracy comes with a time
penalty; AutoProf takes on average 490\,s to produce a profile on an
Intel~Xeon~E5-2650~v3~CPU, a rate of 0.002 galaxies per second.  This rate is
roughly equivalent to the throughput of a human running the
\citet{courteau1996} method.  Pix2Prof processes 0.9 galaxies per second on the
same hardware.  Pix2Prof also offers more flexibility; it can be retrained to
recreate any (semi)~manual galaxy profile fitting pipeline and is therefore not
limited to automation of the \citet{courteau1996} method.

AutoProf is presented in \citet{stone2021}, and its code is available at \url{https://github.com/ConnorStoneAstro/AutoProf}.

\section{Discussion and conclusions} \label{sec.dandc}

While Pix2Prof can rapidly and accurately produce profiles of arbitrary length,
there are some limitations to this technique.  Principally, any profile
produced will be biased to the training set. For instance, if Pix2Prof is
trained on primarily nearby galaxies, it may not yield accurate profiles for
more distant systems whose images will be poorly resolved. Similarly, if the
model is trained on galaxy image--profile pairs as produced by numerical
simulations the model will encode any flaws, incompleteness, or bias inherent
to each simulation and will not encode instrumental effects (e.g.\ read-out
noise) unless properly included.  The same issue will occur if we train on
galaxy image--profile pairs sampled from one survey and deploy the trained
model on a dissimilar survey, for example SDSS, and LSST \citep{sdss, lsst}.
It may be possible to mitigate this problem with an image domain translator
\citep[i.e.][]{cyclegan, pix2pix, stargan} that could transform observations so
that they match a given survey.  Of course, the \citet{courteau1996} measured
profiles may also not entirely reflect the `true' SB profile, due to
modelling assumptions, human bias, and inherent noisiness in measurement.  As
neural networks typically require very large data sets, our relatively small
data set is likely not reflecting the true potential of the model. Therefore, a
larger set of training data could improve the results presented here.
Generating a larger data set from simulated galaxies for training Pix2Prof will
be a future project.

As described in \citet{gLSTM}, due to the vanishing gradient problem a LSTM or
GRU may `forget' an image encoding as it unrolls.  For Pix2Prof, this will
manifest in a loss of accuracy at larger galactocentric radius.  We see this
effect in Fig.~\ref{fig.pix2prof_stats}'s line plot and
Fig.~\ref{fig.channelerrors}, but we cannot disentangle the individual contributions
from image noise and the model architecture.  However, assuming that the noise
is significantly caused by GRU `forgetfulness', future Pix2Prof models could
imitate \citet{gLSTM} and counteract the noise by reinjecting the image
encoding into the GRU's hidden state periodically as it unrolls.  Another
solution could involve adopting an architecture that suffers less from the
vanishing gradient problem, such as the Transformer \citep{AIAYN}.  The
non-sequential nature of a Transformer would also allow us to parallelise
output at inference time, reducing processing time even further.

In Section~\ref{sec.intro} we stressed the need for efficient and fully automated
methods for timely analysis of ultra-large scale astrophysical imaging survey
data.  We believe that Pix2Prof addresses this challenge.  Pix2Prof can predict
any galaxy profile, given the right simulated or observed data set.  Training
Pix2Prof on simulated galaxy images offers additional benefits; the model could
be trained on information that is only inferred indirectly in observations. For
instance, Pix2Prof could train on sets of galaxy image--mass profile pairs
directly in order to predict dark matter halo profiles, as mass profiles cannot
be recovered classically by direct imaging observations.  Furthermore, Pix2Prof
has the potential to automate any galaxy profile fitting routine and be ported
to other forms of galaxy image analysis that may not rely on isophotal
analysis, but still produce a float sequence given a multidimensional array.
These analyses could include galaxy component decompositions, the
characterisation of galaxy interactions and distortions, pixelised stellar
population synthesis, inference of galaxy mass distributions, and more
\citep[e.g.][]{eneev1973, vazdekis2001, GALFIT}.  In a future paper, we will
demonstrate how Pix2Prof can be used to recover simultaneously the galaxy
SB profile as well as the ellipticity profile and curve of
growth of a galaxy. 

An exciting future investigation involves building a system that can predict
properties of unseen classes of objects.  This could be achieved by building up
a `prior' that encodes known objects into a latent space and interpolates
between their latent spatial representations at inference time.  A generative
model like the Generative Adversarial Network \citep[GAN;][]{gan} or Variational
Autoencoder \citep[VAE;][]{vae} could achieve this \citep[i.e.][]{spindler2020}.
Such a model could quickly identify astrophysically `interesting' objects in
a large field survey.  The ability to search for rare objects in large
unstructured data sets will become increasingly more important as new large
scale astronomical surveys come online \citep{pan-starrs, hsc, lsst}.

In summary, we have introduced a fully automated deep learning model for the
extraction of sequential data from galaxy imagery. We have tested this model by
applying it to the specific problem of estimating galaxy SB 
profiles, a process that previously required manual, time-consuming human
intervention.  We have tested our model on unseen galaxy images and found that
our model has an average absolute deviation of 0.34~mag~arcsec$^{-2}$ with an
interquartile range of 0.22~mag~arcsec$^{-2}$, while inferring SB 
profiles over two orders of magnitude faster than the classic
(interactive) algorithm it automates. 

\section*{Data and Code Availability}

The code and trained model used in this paper is available at
\url{https://github.com/Smith42/pix2prof}. The profile data set used to train
the network will be released separately (Arora et al., in preparation).

\section*{Carbon Emissions}

The training of deep learning models requires considerable energy,
contributing to carbon emission and therefore climate change
\citep{energy_ML_model, energy_calculator}. The energy used while training
Pix2Prof on a single NVIDIA V100 GPU is estimated to be $\sim$20~kWh
($5.54~\text{kg}~\text{CO}_2~\text{eq}$) according to the Machine Learning
Impact calculator described in \citet{energy_calculator}.
This is equivalent to driving 28~miles in a typical European passenger
car\footnote{According to the European Environmental Agency:\\
\url{https://www.eea.europa.eu/highlights/average-co2-emissions-from-new-cars-vans-2019}}. 
To counteract further emission from redundant retraining, we follow the
recommendations of \citet{energy_ML_model} and make available the fully trained
model, as well as the code to run it.  Also, we will make available trained
models for any improvements that we make to Pix2Prof in the future.

\section*{Acknowledgements}

This research made use of the University of Hertfordshire's High Performance
Computing facility (\url{http://uhhpc.herts.ac.uk/}).  
We are grateful to the Natural Sciences and Engineering Research Council of Canada, 
the Ontario Government, Queen's University, and the Royal Society for critical support 
through various scholarships and grants.  
M.J.S. thanks Queen's University and the research group of Stéphane Courteau for
their support and hospitality in 2019 while this paper was conceived and developed. 
We also thank reviewers at the NeurIPS~2020 conference, as well as the MNRAS
editor and referee for helpful comments and suggestions. 
The galaxy icon in Fig.~\ref{fig.arch} is by Agata Kuczmi\'nska 
and is available under the \mbox{CC-BY-4.0} licence at
\url{https://goodstuffnononsense.com/hand-drawn-icons/space-icons/}.

\bibliographystyle{mnras}
\bibliography{mkc.bib}

\bsp	
\label{lastpage}
\end{document}